\documentclass[
twocolumn, 10pt,
 amsmath,amssymb,
 aps,
]{revtex4-2}

\usepackage{graphicx}
\usepackage{dcolumn}
\usepackage{bm}
\usepackage{mathtools,bbold}
\usepackage{braket}
\usepackage{xcolor}

\begin{document}

\preprint{APS/123-QED}

\title{Wannier function localisation\\ using Bloch intrinsic atomic orbitals}

\author{Andrew Zhu and David P. Tew}
\affiliation{%
 Physical \& Theoretical Chemistry Laboratory \\ University of Oxford
}%

\date{\today}

\begin{abstract}
We extend the Intrinsic Atomic Orbital (IAO) method for localisation of molecular orbitals to calculate well-localised generalised Wannier functions in crystals using the Pipek--Mezey locality metric. We furthermore present a one-shot diabatic Wannierisation procedure that aligns the phases of the Bloch functions, providing immediate Wannier localisation, which serves as an excellent initial guess for optimisation. We test our Wannier localisation implementation on a number of solid state systems, highlighting the effectiveness of the diabatic preparation, especially for localising core bands. Partial charges of Wannier functions generated using Bloch IAOs align well with chemical intuition, which we demonstrate through the example of adsorption of CO on a MgO surface.
\end{abstract}


\maketitle


\section{Introduction}

Mean field theories, such as Hartree-Fock (HF) or Kohn-Sham density functional theory (DFT)\cite{dft1,dft2}, provide a description of the electronic structure of a system through an one-particle orbital model, which enables interpretation of the bonding in molecules and band structure for materials. However, the canonical molecular orbitals (MOs) or Bloch functions, are typically delocalised across the entire system, and thus do not intuitively map to the interpretation of bonding in terms of overlap of atomic orbitals (AOs), which is a local picture. By applying unitary rotations to the occupied orbitals, one can obtain localised objects, commonly known as Wannier functions (WFs)\cite{Wannier} for periodic systems. Localisation of occupied orbitals aid in interpretation of electronic structure, and also provide a basis for reduced scaling quantum chemistry methods, which exploit this locality to truncate the virtual space\cite{pulay,local1,usyvat1,usyvat2}.

Methods to evaluate localised molecular orbitals have focused upon defining a localisation metric or functional; the stationary  points of this functional thus correspond to localised orbitals. The two most commonly employed metrics are from Foster and Boys (FB)\cite{FB-1,FB-2}, and Pipek and Mezey (PM)\cite{PM}, both of which have been adapted for periodic systems\cite{Marzari-Vanderbilt,Wannier90,MLWFs,Jonsson}.
The FB metric, which minimises the spread of the orbitals, has seen widespread usage, namely through the Wannier90\cite{Wannier90} package, which has now established interfaces with various periodic, plane-wave based, codes\cite{vasp,quantumespresso,castep,abinit}. In contrast, the PM metric, which uses the Mulliken partial atomic charges, is naturally suited to codes employing localised basis sets, under a linear-combination-of-atomic-orbitals (LCAO) framework\cite{CRYSTAL23,pisaniperiodichf1,pisaniperiodichf2,PYSCF,Turbomole}, where AO coefficients are directly accessible and overlaps are easily computed. In addition, WFs localised with the PM metric produces orbitals with separate $\sigma$ and $\pi$ bonding character
, giving advantages in chemical interpretation, as opposed to FB.
The Mulliken charges, however, are ill-defined for non-minimal basis sets, and thus require alternate charge definitions to alleviate this basis set dependence. This issue arises from the near-redundancy of LCAO expansion with large basis sets and is exacerbated in crystals. The intrinsic atomic orbital (IAO) method\cite{knizia2013} is one partial charge estimate that has successfully been applied for molecules.

In this article, we introduce Bloch intrinsic atomic orbitals (Bloch IAOs) as the natural periodic extension of IAOs, and then present the overall optimisation scheme to generate localised WFs with Bloch IAOs. The initial guess is a crucial step in the optimisation and we propose a simple and effective procedure for generating localised orbitals by defining a natural gauge and by constructing diabatic Bloch orbitals and diabatic Wannier functions.
We then present and analyse performance and stability of the optimisation, with particular discussion of the solver's performance when separating core and valence bands. Finally, the chemical interpretability of IBAO WFs is commented upon, using a surface adsorption system as an example.

\section{Theory}
\subsection{Review of Wannier  functions}
Under Born-von-Karman (BvK) boundary conditions, within a linear-combination-of-atomic-orbitals (LCAO) framework, the crystal orbitals are expanded in a basis of Bloch AOs. The Bloch AOs are eigenfunctions of the momentum operator with crystal momentum wave vector $\mathbf{k}$, and are defined as
\begin{equation}\label{Bloch-AO}
	\ket{\mu_{\mathbf{k}}} =\frac{1}{\sqrt{N}}\sum_{\boldsymbol{R}}^N  e^{i \mathbf{k}\cdot \boldsymbol{R}}  \ket{\Tilde{\mu}_{\boldsymbol{R}}}.
\end{equation}
$N$ is the number of unit cells within the BvK `supercell', and $\boldsymbol{R}$ is the lattice vector of the unit cell. $\ket{\Tilde{\mu}_{\boldsymbol{R}}}$ is the infinite sum of real-space AOs from each of the supercells throughout the crystal, and is periodic under the BvK boundary conditions.  The crystal orbitals, also referred to as Bloch functions, are eigenstates of the one-particle Hamiltonian of a periodic system and are given by
\begin{equation}\label{Bloch-1}
	\ket{\psi_{i,\mathbf{k}}} =\sum_{\mu} C_{\mu ,i}^{\mathbf{k}}\ket{\mu_{\mathbf{k}}}.
\end{equation}
The Bloch functions are delocalised across the entire system. By superimposing the Bloch functions of a single band across the first Brillouin zone, a conventional Wannier function\cite{Wannier}, centred on a unit cell given by lattice vector $\boldsymbol{R}$, is given by,
\begin{equation}
	\ket{\phi_{i,\boldsymbol{R}}}=\frac{1}{\sqrt{N}} \sum_{\mathbf{k}}^N e^{-i \mathbf{k}\cdot\boldsymbol{R}} \ket{\psi_{i,\mathbf{k}}}.
\end{equation}
The WFs span the same space as their Bloch counterparts, with translational copies found in each unit cell. 

Bloch functions are defined up to an arbitrary phase only. However, the spatial distribution of the resultant WFs are highly dependant on the relative phases of the contributing Bloch functions. The WFs are thus gauge variant. To obtain localised conventional WFs, the relative phases of the Bloch functions, for a given band, must be optimised. By rotating the gauge such that the Bloch functions appear smooth in reciprocal space, the resulting WFs in real space are in turn localised, as a property of Fourier transforms,

\begin{equation}\label{arbgauge}
	\ket{\phi_{i,\boldsymbol{R}}}=\frac{1}{\sqrt{N}} \sum_{\mathbf{k }}^N e^{-i \mathbf{k}\cdot\boldsymbol{R}} e^{i\theta^{\mathbf{k}}_{i}}\ket{\psi_{i,\mathbf{k}}}.
\end{equation}
A natural gauge for each Bloch function can be defined by requiring that the scalar product of the coefficients between Bloch functions at $\mathbf{k}$ and the $\Gamma$-point $\mathbf{0}$ is real valued. By first computing the phase difference
\begin{equation}
        \sum_{\mu} C_{\mu,i}^{* \mathbf{k}} C_{\mu,i}^{\mathbf{0}}= Re^{i\theta^{\mathbf{k}}_i},
\end{equation}
the Bloch functions can be rotated into their natural gauge $\ket{\psi_{i,\mathbf{k}}^{\mathrm{n}}}$, from their original gauge, $\ket{\psi_{i,\mathbf{k}}^{\mathrm{o}}}$, straightforwardly,
\begin{equation}
    \ket{\psi_{i,\mathbf{k}}^{\mathrm{n}}}=\ket{\psi_{i,\mathbf{k}}^{\mathrm{o}}}
    e^{-i\theta^{\mathbf{k}}_i}.
\end{equation}
For bands that have small dispersion, or minimal mixing, imposing the natural gauge is often sufficient to produce well-localised Wannier functions.

\subsection{Generalised localised Wannier functions}
Generalised Wannier functions are defined by allowing Bloch functions from several bands to mix\cite{Marzari-Vanderbilt},
\begin{equation}\label{fullWF}
	\ket{\phi_{i,\boldsymbol{R}}}=\frac{1}{\sqrt{N}} \sum_{\mathbf{k}}^N \sum_j e^{-i \mathbf{k}\cdot\boldsymbol{R}} U_{j,i}^{\mathbf{k}} \; \ket{\psi_{j,\mathbf{k}}}.
\end{equation}
By allowing mixing between bands, WFs can be further localised not only to a unit cell, but also to atomic sites within a unit cell. The resulting generalised WFs are even more strongly `non-unique' than the conventional definition, and their locality is highly dependant on the choice of $ U_{ji}^{\mathbf{k}}$. To ensure real-valued WFs, inversion symmetry about the $\Gamma$-point must be imposed between the Bloch functions, $\ket{\psi_{i,\mathbf{k}}}=(\ket{\psi_{i,-\mathbf{k}}})^*$.
The Bloch functions at the $\Gamma$-point are real, following convention, and the choice of unitary carries the constraint of $\boldsymbol{U}^{\mathbf{k}}=(\boldsymbol{U}^{\mathbf{-k}})^*$.

Localisation of WFs is achieved by varying $\boldsymbol{U}^{\mathbf{k}}$ to optimise a chosen locality metric. The Foster-Boys\cite{FB-1,FB-2} and Pipek-Mezey\cite{PM} metrics are two of the most important examples. The FB method localises orbitals by defining a metric which minimises the orbital spread, as given by its variance,
\begin{equation}{\label{FB}}
\braket{O}_{\textbf{FB}}=\sum_{i}\bra{\phi_{i,\mathbf{0}}}\mathbf{r^2}\ket{\phi_{i,\mathbf{0}}}-\bra{\phi_{i,\mathbf{0}}}\mathbf{r}\ket{\phi_{i,\mathbf{0}}}^2.
\end{equation} 
Marzari and Vanderbilt\cite{Marzari-Vanderbilt} generalised the FB approach, originally conceived for molecules, to evaluate localised WFs, creating so called Foster-Boys Wannier functions (FBWFs). The Wannier90 package\cite{Wannier90}, which employs this method, has been widely used amongst the solid-state community\cite{MLWFs}.

The original Pipek-Mezey metric was defined as the sum of squares of the Mulliken partial charges\cite{PM}. J\'{o}nsson et al.\cite{Jonsson} introduced a scheme to generate localised WFs using the PM metric (PMWFs),
\begin{equation}{\label{PMmetric}}
\braket{O}_{\textbf{PM}}=\sum_{\boldsymbol{R},A,i}|Q_i^{A_{\boldsymbol{R}}}|^p=\sum_{\boldsymbol{R},A,i}\bra{\phi_{i,\mathbf{0}}}\hat{P}_{A_{\boldsymbol{R}}}\ket{\phi_{i,\mathbf{0}}}^p, 
\end{equation}
where  $Q_i^{A_{\boldsymbol{R}}}$ is the Mulliken charge\cite{PM}
associated with WF $i$ on atom $A$, situated in unit cell $\boldsymbol{R}$, evaluated using the WFs located in the reference unit cell. $\hat{P}_{A_{\boldsymbol{R}}}$ projects onto a basis of AOs  centred on atom $A_{\boldsymbol{R}}$, given by,
\begin{equation}
   \label{eq:a-projector}
	\hat{P}_{A_{\boldsymbol{R}}}=\sum_{\mu \in A} \sum_{\nu,\boldsymbol{R'}} \ket{\Tilde{\mu}_{\boldsymbol{R}}} (S^{-1})_{\mu,\boldsymbol{R}}^{\nu,\boldsymbol{R'}}\bra{\Tilde{\nu}_{\boldsymbol{R'}}},
\end{equation}
where $S_{\mu,\boldsymbol{R}}^{\nu,\boldsymbol{R'}}=\braket{\Tilde{\mu}_{\boldsymbol{R}}|\Tilde{\nu}_{\boldsymbol{R'}}}$ is the overlap. Under a LCAO approach, where atomic orbital coefficients are naturally available, computing the PM metric is extremely straightforward, as opposed to the FB metric, which has been used more commonly with plane wave basis sets. In addition, the strong interpretability of PMWFs, producing orbitals with $\sigma$ and $\pi$ separation\cite{Jonsson}, further motivates our choice to utilise this metric, as opposed to the `banana' bonds found within the FBWF scheme.

The  value of the penalty exponent, $p$, is typically $2$, or $4$. If $p$ is chosen to equal $1$, then Eq.\ref{PMmetric} reduces to the normalisation criteria of the bands,
\begin{equation}{\label{normalisation}}
\sum_{\boldsymbol{R},A}\sum_{i}^{\mathrm{n_{occ}}}Q_i^{A_{\boldsymbol{R}}}=\mathrm{n_{occ}},
\end{equation}
where $\mathrm{n_{occ}}$ are the number of occupied bands.

A key issue with the original definition of the PM metric is that the Mulliken partial charges do not possess a complete basis set limit, meaning they are undefined for non-minimal basis sets. Alternative charge definitions for MOs have been suggested\cite{Lehtola-partial-charge,Cioslowski-charge-estimate,Alcoba-charge-estimate}, 
that alleviate this basis set dependence. Lehtola and J\'{o}nsson demonstrated that the localised orbitals obtained were largely independent of the chosen partial charge estimate\cite{Lehtola-partial-charge}, providing significant freedom in choice. Other alternative charge definitions have been implemented for periodic systems, such as real space partitioning of orbital charge densities\cite{Jonsson,bader-charge-grid,hirshfeld-charge-periodic}, or other types of projection onto a pre-determined set of minimal basis functions\cite{Clement}.

The intrinsic atomic orbital method (IAO), as proposed by Knizia \cite{knizia2013}, is one choice of an alternative partial charge estimate that has been employed successfully for molecules. Using a free-atom minimal basis as a template, contraction coefficients from the original basis to IAOs are defined such that the occupied orbitals are exactly represented, which provides a consistent assignment of charge to atomic centres. Localised MOs using IAOs align well with chemical intuition, and quantitative measures such as partial charges and populations are shown to be resistant to changes in original basis, and are consistent with chemical understanding, leading to the method being implemented in many quantum chemistry packages\cite{Turbomole,PYSCF,molpro,dirac}. We thus propose to adapt IAOs to construct a charge metric suitable to localise Wannier functions, $Q_i^{A_{\boldsymbol{R}}^{\mathrm{IAO}}}$, using Bloch intrinsic atomic orbitals (Bloch IAOs).

\subsection{Bloch intrinsic atomic orbitals}
Given the success of IAOs within molecular schemes, we believe a $\mathbf{k}$ space extension to periodic systems would be desirable. Sch\"{a}fer et al.\cite{Gruneis} demonstrate the use of IAOs to evaluate localised Wannier functions for a $\Gamma$-point-only calculation, following the molecular formulation as described by Janowski\cite{janowski2014near}. Cui et al.\cite{Cui-k-IAOs} construct crystal IAOs, from which projected AOs are evaluated. We employ similar principles in our generalisation to $\mathbf{k}$ space, but crucially outline the additional augmentations needed to construct localised WFs, optimised using the PM metric, as a full periodic adaption of the IAO method.

We choose to adapt Knizia's method\cite{knizia2013}, such that a set of intrinsic atomic orbitals are constructed for each $\mathbf{k}$ point within our Monkhorst-Pack quadrature mesh\cite{kmesh}. 
The Bloch IAOs are able to exactly describe the original occupied Bloch functions, providing a basis independent charge metric for WFs.

The original Bloch functions (Eq.\ref{Bloch-1}) are expressed in terms of Bloch AOs in the original basis set, labelled $B_1$. Analogous to Knizia's approach, a minimal basis, $B_2$, of free-atom AOs is first chosen, from which corresponding Bloch AOs are obtained, $\ket{\rho_{\mathbf{k}}}$ where $\rho \in B_2$, from Eq.\ref{Bloch-AO}. 

The following projection operators are defined,
\begin{equation} \label{20-03_2}
	\hat{P}_{12}^{\mathbf{k}}=  \sum_{\substack{\mu,\nu\in\mathrm{B_1}}} \ket{\mu_{\mathbf{k}}} (S^{-1}_{1})_{\mu,\nu}^{\mathbf{k}} \bra{\nu_{\mathbf{k}}}
\end{equation}

\begin{equation} \label{20-03_3}
	\hat{P}_{21}^{\mathbf{k}}=  \sum_{\substack{\rho,\sigma\in\mathrm{B_2}}} \ket{\rho_{\mathbf{k}}} (S^{-1}_{2})_{\rho,\sigma}^{\mathbf{k}}\bra{\sigma_{\mathbf{k}}},
\end{equation}
where $(S_{1})_{\mu,\nu}^{\mathbf{k}} = \langle \mu_\mathbf{k} \vert \nu_\mathbf{k} \rangle$ and $(S_{2})_{\rho,\sigma}^{\mathbf{k}} = \langle \rho_\mathbf{k} \vert \sigma_\mathbf{k} \rangle$ are the Bloch-AO overlap matrices in the original and minimal basis sets, respectively. Using these operators, depolarisied occupied Bloch functions are obtained through
\begin{equation} \label{depolarise}
		\{\ket{\tilde{\psi}_{i,\mathbf{k}}}\}=\mathrm{orth}\{ \hat{P}_{12}^{\mathbf{k}}\hat{P}_{21}^{\mathbf{k}}\ket{\psi_{i,\mathbf{k}}}\}
\end{equation}
or in matrix form:
\begin{equation}\label{depolarise-matrix}
	\mathbf{\tilde{C}^k}=\mathrm{orth}\{\mathbf{P_{12}^kP_{21}^kC^k}\},
\end{equation}
Here $\mathrm{orth}\{\}$ denotes symmetric orthogonalisation and the transfer matrices are $\mathbf{P_{12}^k}=(\mathbf{S}_1^{-1})^\mathbf{k} \mathbf{S}_{12}^\mathbf{k}$ and $\mathbf{P_{21}^k}=(\mathbf{S}_2^{-1})^\mathbf{k} \mathbf{S}_{21}^\mathbf{k}$, where $(S_{12})^\mathbf{k}_{\mu \rho}= \langle \mu_\mathbf{k} \vert \rho_\mathbf{k} \rangle$ and $\mathbf{S}_{21}^\mathbf{k} = \mathbf{S}_{12}^{\mathbf{k}\dagger}$. The projector onto the depolarised occupied Bloch functions is $\tilde{O}^{\mathbf{k}}=\sum_{i}\ket{\tilde{\psi}_{i,\mathbf{k}}}\bra{\tilde{\psi}_{i,\mathbf{k}}}$.

The Bloch IAOs are the minimal Bloch AO basis that contains both the depolarised and polarisation contributions and are defined through
\begin{equation} \label{14-03_3}
	\ket{\rho_{\mathbf{k}}^{\mathrm{IAO}}}= (O^{\mathbf{k}}\tilde{O}^{\mathbf{k}}+(1-O^{\mathbf{k}})(1-\tilde{O}^{\mathbf{k}}))\hat{P}^{\mathbf{k}}_{12}\ket{\rho_{\mathbf{k}}}.
\end{equation}
In matrix notation, Eq.\ref{14-03_3} is given as:
\begin{equation} \label{25-9}
	\begin{split}
		\mathbf{A^k}= &\mathbf{C^{\mathbf{k}}C^{k\dagger}S_1^k \tilde{C}^k \tilde{C}^{k\dagger} S_{12}^k} +\\
		&\mathbf{(1-C^{\mathbf{k}}C^{k\dagger}S_1^k )(P_{12}^k-\tilde{C}^k \tilde{C}^{k\dagger}S_{12}^k )},
	\end{split}
\end{equation}
where $\mathbf{A^k}$ are the contraction coefficients from Bloch AOs to Bloch IAOs at each $\mathbf{k}$ point, and $\mathbf{1}$ is the identity in the space of $B_1$. Janowski\cite{janowski2014near} and Knizia\cite{knizia2013} both outline a simpler definition for the IAOs, which is equivalent under the assumption that $B_2$ can be directly expressed in $B_1$. Having implemented both schemes, we note that the output Bloch IAOs are very similar, with no significant difference in localisation performance. Finally, the coefficients of the occupied Bloch functions in the Bloch IAO basis are given by 

\begin{equation} \label{BlochcoeffIAO}
		\mathbf{C^{k(\mathrm{IAO})}= (S^{k(\mathrm{IAO})})^{-1} A^{k\dagger} S_1^k C^k},
\end{equation}
\begin{equation} 
	\mathbf{S^{k(\mathrm{IAO})}=A^{k\dagger} S_1^k A^k} .
\end{equation}
In the original molecular implementation, the output IAO coefficients are symmetrically orthogonalised. However, in the periodic case, we choose not to do so. The orthogonalisation procedure introduces arbitrary phases to the Bloch functions in the IAO basis, specifically when obtaining the eigenvectors of the IAO coefficient matrix. The relative phase differences between $\mathbf{k}$ points are thus altered compared to the original Bloch functions expressed in $B_1$, leading to issues when optimising the set of unitary matrices, across the Brillouin zone, in the IAO basis, since they do not correspond to the original Bloch functions. The simplest solution to remove this additional gauge problem is to leave the IAOs un-orthogonalised.
The `depolarised' Bloch functions, given by Eqs.\ref{depolarise} and \ref{depolarise-matrix}, which are orthogonalised, avoid this issue, because any phase augmentation is cancelled in the projector $\tilde{O}^k$. 

In summary, obtaining the Bloch IAOs is numerically straightforward, requiring only a free atom basis, and its corresponding Bloch AO overlaps to perform the matrix multiplication steps. Computation of inverse overlap matrices can be avoided by solving instead with a Cholesky decomposition. The Bloch coefficients in the IAO basis and the IAO overlap matrix can then be used in a Pipek-Mezey style optimisation to obtain optimally local Wannier functions. The PM projector (Eq.\ref{eq:a-projector}), in the Bloch IAO basis, is now defined as 
\begin{equation}
   \label{eq:a-projector-iao}
	\hat{P}_{A_{\boldsymbol{R}}}^{\mathrm{IAO}}=\sum_{\rho \in A} \sum_{\sigma,\boldsymbol{R'}} \ket{\Tilde{\rho}^{\mathrm{IAO}}_{\boldsymbol{R}}} (S^{-1})_{\rho,\boldsymbol{R}}^{\sigma,\boldsymbol{R'}(\mathrm{IAO})}\bra{\Tilde{\sigma}^{\mathrm{IAO}}_{\boldsymbol{R'}}},
\end{equation} where $S_{\rho,\boldsymbol{R}}^{\sigma,\boldsymbol{R'}(\mathrm{IAO})}=\braket{\Tilde{\rho}^{\mathrm{IAO}}_{\boldsymbol{R}}|\Tilde{\sigma}_{\boldsymbol{R'}}^{\mathrm{IAO}}}$ is the IAO overlap in real space, obtained from Fourier transforming $\mathbf{S^{k(\mathrm{IAO})}}$.

\subsection{Localisation procedure}
Recent work obtaining PM localised WFs and MOs have involved global optimisation algorithms to determine the stationary points of the functional\cite{Clement,Jonsson,Lehtola,Lehtola-partial-charge}.
Clement et al.\cite{Clement} recently demonstrated that a solver using the Broyden-Fletcher-Goldfarb-Shanno (BFGS) algorithm leads to significantly faster convergence compared to previous steepest ascent (SA) or conjugate gradient implementations.
Our localisation procedure uses a BFGS based algorithm, that we employ in conjunction with our Bloch IAO charges. To generate an effective initial guess for the optimisation, a novel procedure generates approximately localised WFs, which we call diabatic Wannier functions.

\subsubsection{Diabatic Wannierisation}
The initial guess for the WFs is an important step in the localisation procedure in order to avoid encountering local maxima. Methods which project Bloch functions onto a set of trial functions have been outlined\cite{Marzari-Vanderbilt,MLWFs}, whilst other implementations ensure the unitary space is probed fully by running multiple calculations using randomly sampled unitary matrices\cite{Jonsson,Clement}. Clement et al.\cite{Clement} combine random unitary sampling with a procedure to remove the gauge freedom of the Bloch functions.

As mentioned prior, Bloch functions are defined with an arbitrary gauge (Eq.\ref{arbgauge}). By fixing the gauge such that the variations between Bloch functions in $\mathbf{k}$-space are gradual, the Fourier transform produces WFs which are largely localised to a single cell, serving as an excellent starting guess for further optimisation. We have defined the natural gauge to be where the scalar product of the coefficients between Bloch functions within a band at $\mathbf{k}$ and the $\Gamma$-point $\mathbf{0}$ is real. For generalised WFs (Eq.\ref{fullWF}), where the gauge uncertainty is increased by mixing bands, we extend the intuition of the natural gauge, to construct diabatic Bloch orbitals and diabatic Wannier functions. First, the Bloch orbitals of the $\Gamma$-point are localised by orthogonal transformation,
\begin{align}
	\vert \psi_{i,\mathbf{0}}  \rangle = \sum_{j} \vert \psi_{j,\mathbf{0}} \rangle O_{ji}.
\end{align}
The Bloch orbitals of the remaining $\mathbf{k}$-points are then chosen to be those with maximal similarity with the $\Gamma$-point. The locality of the orbitals of the $\Gamma$-point is thus transferred diabatically across
the first Brillouin zone. This is obtained by calculating the unitary matrices, outside the $\Gamma$-point, which give the minimal least squares difference to the Bloch coefficients of the $\Gamma$-point,
\begin{equation}\label{initial-guess}
	\underset{\boldsymbol{U}_{\mathbf{k}}}{\mathrm{min}}||C_{\mu ,p}^{0}- \sum_j C_{\mu ,j}^{\mathbf{k}} U^{\mathbf{k}}_{j,p}||^2.
\end{equation}
where $||\dots||$ is the Frobenius\cite{matrix-analysis} norm. As this is an example of an Orthogonal Procrustes problem\cite{procrustes}, a solution can be easily obtained via the singular value decomposition of the product of Bloch coefficients $\boldsymbol{C^{\mathbf{k}\dagger}C^{0}}$, 
\begin{align}
	\vert \psi_{i,\mathbf{k}} \rangle &= \sum_{\mu,j,j'} C_{\mu,j}^{\mathbf{k}}U_{j,j'} V_{j',i} \vert \mu_\mathbf{k} \rangle \\
	\boldsymbol{C^{\mathbf{k}\dagger}C^{0}}&= \mathbf{U} \mathbf{\Sigma} \mathbf{V} .
\end{align}
In this work, we employ a convenient approximate localisation procedure for the Bloch orbitals of the $\Gamma$-point, where we simply replace them with the Cholesky vectors of the $\Gamma$-point density, ensuring computation of the diabatic WFs is fast.

\subsubsection{Optimisation of the PM metric}
We implement a gradient based optimisation method to obtain the stationary points of the PM functional. Similar to Clement et al.\cite{Clement} and Lehtola and J\'{o}nsson\cite{Lehtola},
a Riemannian geometry approach is adopted to maintain the unitary constraint, as outlined in ref. \cite{AbrudanCG,Abrudan}. This method has proved successful since the unitary constraint is maintained implicitly, whilst other methods, including Lagrange multipliers\cite{ica}, may suffer from slow convergence or only obtain a solution which only approximately maintains orthonormality.

Given the extensive discussion of the unitary optimisation algorithm in ref. \cite{Abrudan,AbrudanCG}, we only briefly outline our procedure here. Crucially, the PM charge metric and all associated expressions are evaluated in the Bloch IAO basis, using $\mathbf{C^{k(\mathrm{IAO})}}$ (Eq.\ref{BlochcoeffIAO}). The real space IAO overlaps, $S_{\sigma,\boldsymbol{R'}}^{\rho,\boldsymbol{R}(\mathrm{IAO})}$, are also used.

In the following expressions, the IAO labels are omitted for clarity. 
The Bloch IAO charges are defined as
\begin{equation}
	\begin{split}
&Q_i^{A_{\boldsymbol{R}}^{\mathrm{IAO}}}=\bra{\phi_{i,\mathbf{0}}}\hat{P}_{A_{\boldsymbol{R}}}^{\mathrm{IAO}}\ket{\phi_{i,\mathbf{0}}}= \\
	&\frac{1}{N^2}\sum_{\rho \in A} \Bigr[\sum_{\substack{j,\mathbf{k}}}
	\bar{C}_{\rho,j}^{*\mathbf{k},\boldsymbol{R}} \; U_{j,i}^{*\mathbf{k}} \;\Bigr] \Bigr[\sum_{j',\mathbf{k'}} C_{\rho,j'}^{\mathbf{k'}, \boldsymbol{R}} \; U_{j',i}^{\mathbf{k'}} \Bigr],
	\end{split}
\end{equation}
where we have introduced $C_{\rho,j}^{\mathbf{k},\boldsymbol{R}}=C_{\rho,j}^{\mathbf{k}} e^{i\mathbf{k}\boldsymbol{R}}$ and $\bar{C}_{\rho,j}^{\mathbf{k},\boldsymbol{R}} = \sum_{\sigma,\boldsymbol{R'}} C_{\sigma,j}^{\mathbf{k},\boldsymbol{R'}} S_{\sigma,\boldsymbol{R'}}^{\rho,\boldsymbol{R}}$. The Euclidean derivative of the PM functional, $\braket{O}_{\textbf{PM}}$, with respect to the unitary at $\mathbf{k}$, is given by:

\begin{equation}\label{pm-gradient}
	\begin{split}
		&\frac{\partial \braket{O}_{\textbf{PM}}}{\partial U_{j,i}^{*\mathbf{k}}}= \frac{\partial \braket{O}_{\textbf{PM}}}{\partial Q_i^{A_{\boldsymbol{R}}^{\mathrm{IAO}}}} \frac{\partial Q_i^{A_{\boldsymbol{R}}^{\mathrm{IAO}}}}{\partial U_{j,i}^{*\mathbf{k}}} \\
		&=\frac{p}{N^2} \sum_{\boldsymbol{R},A} |Q_i^{A_{\boldsymbol{R}}^{\mathrm{IAO}}}|^{p-1} \sum_{\rho \in A} \bar{C}_{\rho,j}^{*\mathbf{k},\boldsymbol{R}}\Bigr[ \sum_{j',\mathbf{k'}}C_{\rho,j'}^{\mathbf{k'},\boldsymbol{R}} U_{j',i}^{\mathbf{k'}}\Bigr]\\
		&+ C_{\rho,j}^{*\mathbf{k},\boldsymbol{R}} \Bigr[\sum_{\substack{j'', \mathbf{k''}}} \bar{C}_{\rho,j''}^{\mathbf{k''},\boldsymbol{R}} \; U_{j'',i}^{\mathbf{k''}}  \Bigr].
	\end{split}
\end{equation}

 The Riemannian gradient, $\boldsymbol{G}_{\mathbf{k}}$, can then be transformed from the Euclidean gradient, $\boldsymbol{\Gamma}_{\mathbf{k}}=\frac{\partial \braket{O}_{\textbf{PM}}}{\partial \boldsymbol{U}^*_{\mathbf{k}}}|_{\boldsymbol{U}_{\mathbf{k}}}$, by
\begin{equation}
	\boldsymbol{G}_{\mathbf{k}}=\boldsymbol{\Gamma}_{\mathbf{k}}(\boldsymbol{U}_{\mathbf{k}})^{\dag} -\boldsymbol{U}_{\mathbf{k}}(\boldsymbol{\Gamma}_{\mathbf{k}})^{\dag}.
\end{equation}
The `two loop recursion' version of the limited-memory BFGS algorithm (l-BFGS)\cite{Nocedal} is used, as first implemented for WFs by Clement et al.\cite{Clement}, to obtain a search direction, $\{\boldsymbol{H}_{\mathbf{k}}\}$. The matrix elements in the upper triangle of the anti-Hermitian matrices for the Riemannian gradient, $\{\boldsymbol{G}_{\mathbf{k}}\}$, form the gradient vector for the l-BFGS algorithm, ensuring the output search direction is located on the unitary manifold.
Given the requirement for the output WFs to be real, as mentioned earlier, our gradient vector is comprised of a total $\frac{N-1}{2}o^2 +\frac{o(o-1)}{2}$ real numbers, where $o$ is the number of Bloch functions being localised. 

To obtain a suitable step size, $\mu_{\mathrm{opt}}$, an Armijo\cite{Abrudan,polak-optimisation}
and Wolfe line search\cite{Nocedal} 
were both implemented. If the line search along the l-BFGS direction fails, the search direction is reset to the steepest ascent vector, and the line search is then repeated.

The unitary matrices at each $\mathbf{k}$ point are updated,
\begin{equation}
	\boldsymbol{U}_{\mathbf{k},\mathrm{new}}= \;e^{\mu_{\mathrm{opt}}\boldsymbol{H}_{\mathbf{k}}}\boldsymbol{U}_{\mathbf{k},\mathrm{old}},
\end{equation}
until the norm of the Riemannian gradient decreases below a threshold. The output unitary can then be applied directly to the original Bloch functions, to obtain localised WFs in the $B_1$ basis, using Eq.\ref{fullWF}.

\section{Results}
\subsection{Computational Details}
The Bloch IAO procedure and the Pipek-Mezey Wannier function localisation have been implemented in a developmental version of the TURBOMOLE\cite{Turbomole} package.
The initial mean-field Bloch functions were obtained through the periodic Hartree-Fock procedure, within the \verb:riper: module of the TURBOMOLE package\cite{riper1,riper2,riper3,riper4,riper5}. To generate the IAOs, a minimal basis was constructed from HF calculations of isolated atoms in the cc-pVTZ\cite{ccpvtz-b-ne,ccpvtz-b-ne,ccpvtz-mg} basis, as already implemented within TURBOMOLE to construct molecular IAOs.
The PM functional was evaluated with a penalty exponent of $p=4$, rather than $2$, as shown in Eq.\ref{PMmetric}, due to better localisation for $\pi$ character orbitals, as discussed in prior works\cite{knizia2013,Lehtola-partial-charge}.
 
An Armijo step size method and a Wolfe line search were tested in the localisation procedure. It is known that a line search fulfilling the Wolfe conditions ensures stability of the BFGS updates, by ensuring the approximate Hessian, within our maximisation problem, is negative definite\cite{Nocedal}.
However, computation of the line search is costly, since multiple gradient evaluations are required along the trial direction. By contrast, the Armijo line search only requires PM metric values to be evaluated, and we observed its convergence performance to be similar to the Wolfe line search, with shorter wall times. The Armijo search was employed in all calculations subsequently discussed in this article.

\begin{table}
	\caption{\label{tab:table1}
	Insulating and semiconducting systems used to test the Bloch IAO and PM localisation procedure }
\begin{ruledtabular}
	\begin{tabular}{cccccccc}
		&System&Basis $B_1$& Monkhorst-Pack mesh size\\
		\hline
		&diamond&pob-TZVP\cite{pob-TZVP}&11,11,11\\
		&silicon&pob-TZVP&11,11,11\\
		&boron nitride&pob-TZVP&15,15\\
		&graphene&pob-TZVP&15,15\\
		&MgO&def-SVP&11,11,11\\
		&SiO2&pob-TZVP&5,5,5\\
		&\emph{trans}-$\mathrm{(C_2H_2)}_{\infty}$&pob-TZVP&101\\
		&(4,4) C-nanotube&def2-SVP&11&&&&\\

	\end{tabular}
\end{ruledtabular}
\end{table}

Table \ref{tab:table1} details the insulating and semi-conducting systems used to probe the performance of the IAO PM localisation scheme. The original basis set, $B_1$, used in the mean field calculation, and the Monkhorst-Pack\cite{kmesh} mesh, are shown.
All test systems utilised the universal Coulomb-fitting auxiliary basis sets\cite{auxbas}
with the exception of the magnesium oxide and carbon nanotube systems, which employed auxiliary functions optimised for the def-SVP and def2-SVP basis sets, respectively\cite{auxbas}. Unit cell parameters and geometries are provided in the supplementary information.

\subsection{Overall performance}
\begin{figure*}
	\includegraphics[scale=0.7]{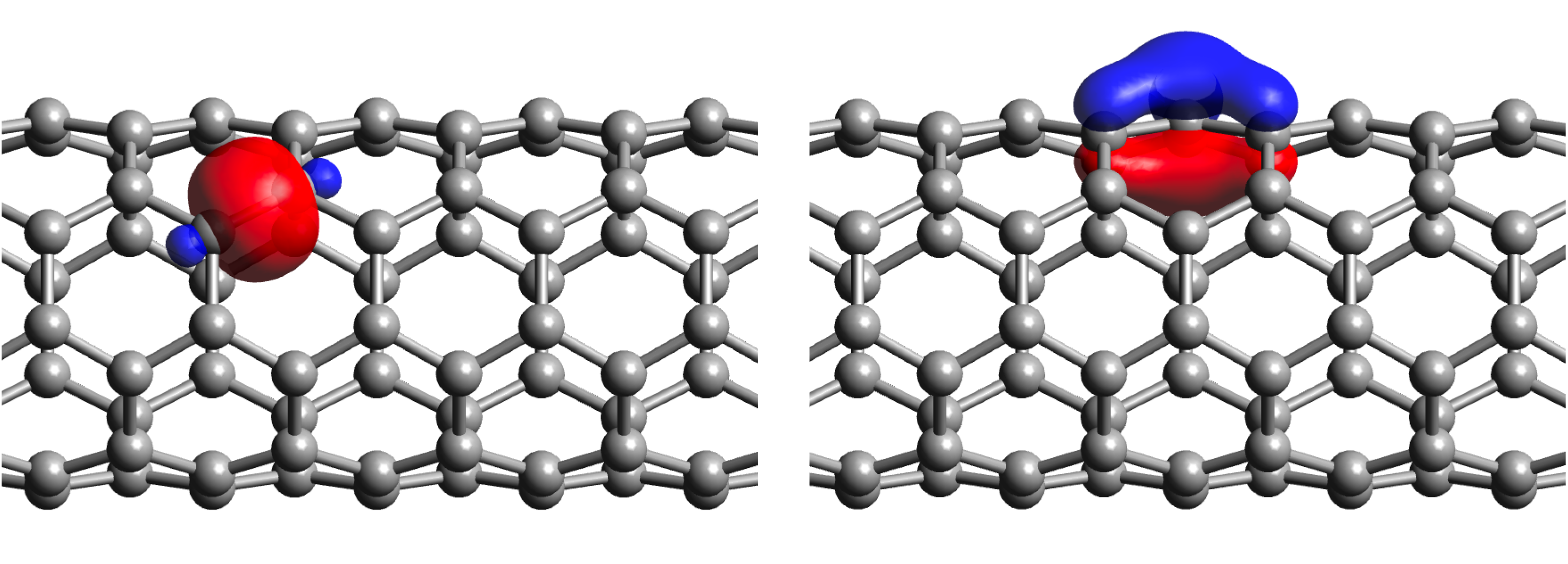}
	\caption{\label{fig:nanotube}Bloch IAO localised Wannier functions of (4,4) nanotube, showing $\sigma$ (left) and $\pi$ (right) bonding character. An isosurface value of 0.05 was used.}
\end{figure*}

\begin{table*}
	\caption{\label{tab:table2}
	PM metric values of WFs from the initial SCF calculation, after rotation into the natural gauge, after diabatic Wannierisation and after PM optimisation. Number of iterations to converge PM metric, after initial diabatic preparation, using l-BFGS or steepest ascent.} 
	\begin{ruledtabular}
		\begin{tabular}{cccccccc}
            &System&$\braket{O}_{\textbf{PM}}$(SCF)&$\braket{O}_{\textbf{PM}}$(Nat.)&$\braket{O}_{\textbf{PM}}$(Dia.)&$\braket{O}_{\textbf{PM}}$ (Opt.)&l-BFGS& SA\\
			\hline
			&diamond&$1.49 \times 10^{-8}$&$6.13 \times 10^{-2}$&1.92&2.66&35&100\\
			&silicon&$2.40 \times 10^{-2}$&$6.64 \times 10^{-2}$&9.79&10.66&42&114\\
			&boron nitride&$1.20 \times 10^{-5}$&2.36&3.19&3.37&55&731\\
			&graphene&$3.37 \times 10^{-6}$&$8.37 \times 10^{-2}$&1.94&2.56&42&60\\
			&MgO&$2.88 \times 10^{-1}$&3.87&9.56&9.61&8&744\\
			&SiO2&$5.96 \times 10^{-4}$&9.55&30.80&37.29&338&1794\\
			&\emph{trans}-$\mathrm{(C_2H_2)}_{\infty}$&$4.92 \times 10^{-5}$&$4.25 \times 10^{-2}$&2.17&2.76&43&1517\\
			&(4,4) C-nanotube&$1.65 \times 10^{-3}$&$4.34 \times 10^{-3}$&33.70&40.60&62&261\\
		\end{tabular}
	\end{ruledtabular}
\end{table*}

Table \ref{tab:table2} reports the performance of the Bloch IAO localisation scheme. The values of the PM metric are presented for the WFs of the SCF calculation, after rotation into the natural gauge, after diabatic Wannierisation and after PM optimisation. The number of iterations required to localise the PM objective function with l-BFGS, compared to steepest ascent (SA), using diabatic WFs as the initial guess, are also given. The convergence threshold for the PM gradient norm was set to $10^{-5}$, with the exception of boron nitride (BN), which was set to $10^{-6}$. All occupied orbitals are included in the optimisation.

In all cases, the final PM values from the l-BFGS optimisation and steepest ascent were equal, to a threshold of $10^{-5}$ (or $10^{-6}$ for BN), confirming that the output WFs were equally localised. A tighter threshold was required for BN in order for the output WFs from l-BFGS and SA to agree to target precision. As demonstrated by Clement et al.\cite{Clement}, we confirm that utilising l-BFGS, compared to SA, markedly improves convergence performance. In some systems, a tenfold reduction in iterations to converge is observed, if not greater. The PM optimiser performs robustly across the range of insulating and semiconducting materials explored, successfully localising every test system. With the exception of silicon dioxide, all test systems converge within 100 iterations with l-BFGS. 
A larger number of iterations was required for silicon dioxide. We observed that the step size was very small, which indicates that the Hessian description of the landscape in this case may be poor. Despite this, l-BFGS still converges five times faster compared to SA, for this example, showing the robustness of the scheme.

We stress that using diabatic WFs as the initial preparation has an important role in the robustness and quality of the final localised WFs. As seen in Table \ref{tab:table2}, the PM values after diabatic Wannierisation are remarkably close to the final optimised values, showing that a significant degree of locality has been captured through the diabatisation. Our experiments using randomly generated unitary matrices as the initial guess led to final WFs with PM values that consistently were smaller than that obtained from the diabatic preparation, and never greater. The choice of objective functional, and the parameterisation employed for the gradient, gives an optimisation landscape with many local maxima and use of an appropriate initial guess, such as diabatic WFs, is required to ensure the global maxima is located. Direct comparison of the number of iterations required to converge, using a random guess and the diabatic preparation, is often not possible since the final WFs are usually inequivalent. Although we have not verified it in this work, we predict that localising the Bloch functions of the $\Gamma$-point with an IAO procedure instead of via Choleski decomposition would further increase $\braket{O}_{\textbf{PM}}$(Dia.) and would reduce the number of iterations required for full optimisation.

The initial PM values from straight Wannierisation of the SCF Bloch functions are very small. It should be noted that due to the different gauges of the Bloch functions in separate calculations, the values for $\braket{O}_{\textbf{PM}}$(SCF) can vary arbitrarily. The values reported in Table \ref{tab:table2} are in fact the largest value of $\braket{O}_{\textbf{PM}}$(SCF) taken from 10 separate calculations. Throughout our testing, we have never observed any example where $\braket{O}_{\textbf{PM}}$(SCF) has been greater or similar in magnitude to $\braket{O}_{\textbf{PM}}$(Dia.). Applying the natural gauge to the Bloch functions increases the PM values in all cases, confirming that the natural gauge smooths the Bloch functions in reciprocal space. For crystals with well-separated bands formed from weakly interacting AOs, simply applying the natural gauge results in well-localised WFs. In all cases, $\braket{O}_{\textbf{PM}}$(Nat.)) sit in between the SCF and diabatic values. 

Figure \ref{fig:nanotube} presents example localised WFs of the (4,4) nanotube system, described in Table \ref{tab:table1}. All figures were plotted using Avogadro\cite{Avogadro}.
The left subfigure clearly shows a carbon-carbon $\sigma$ bond, whilst the right subfigure illustrates a $\pi$ character WF. This $\pi$ and $\sigma$ separation is observed in the other test systems, demonstrating that the Bloch IAO localisation procedure retains the advantages of the original PM metric.

\subsection{Separate optimisation of valence and core bands}
It is often desirable to localise WFs for the core and valence bands separately. For example, within local correlation theories for molecules, obtaining localised occupied orbitals with no contribution from uncorrelated core orbitals is necessary to compute accurate correlation energies\cite{pnof12}. 
In view of this, we report that the Bloch IAO localisation scheme performs exceptionally well in this scenario, due to the quality of the diabatic WFs, which almost immediately generates localised core orbitals. 

\begin{table}
	\caption{\label{tab:table3}}
	Number of iterations to converge PM metric with initial diabatic preparation, with core and valence band separation 
	\begin{ruledtabular}
		\begin{tabular}{cccccccc}
			&System&Initial core PM gradient&core&valence\\
			\hline
			&diamond&$1.08 \times 10^{-4}$&2&20\\
			&silicon&$8.73 \times 10^{-4}$&3&24\\
			&boron nitride&$9.70 \times 10^{-4}$&3&27\\
			&graphene&$1.91 \times 10^{-4}$&2&22\\
			&MgO&$5.17 \times 10^{-5}$&2&4\\
			&SiO2&$6.55 \times 10^{-3}$&2&113\\
			&\emph{trans}-$\mathrm{(C_2H_2)}_{\infty}$&$1.67 \times 10^{-4}$&3&22\\
			&(4,4) C-nanotube&$6.81 \times 10^{-2}$&3&29&&&\\
		\end{tabular}
	\end{ruledtabular}
\end{table}

Table \ref{tab:table3} shows the norm of the PM gradient for the core bands after the diabatic initial preparation, and the subsequent iterations required to localise with the PM metric, using l-BFGS, for both the core and valence bands, across all the test systems. The PM gradient norms of our initial core WFs are all already close to the convergence threshold ($10^{-5}$, or $10^{-6}$ for BN), and localise within 3 l-BFGS iterations, demonstrating that the diabatic preparation yields nearly optimally local core WFs. The number of iterations required to localise the core increases typically only by 1 or 2 iterations when the convergence threshold is increased to $10^{-8}$. 
Across all test systems, the difference between the final localised PM values for the core orbitals and $\mathrm{n_{core}}$ was within $3\times10^{-2}$, where $\mathrm{n_{core}}$ is given by Eq.\ref{normalisation}, summing only over core bands. This indicates that the core orbitals are localised to the global maxima, given that $\mathrm{n_{core}}$ is the upper bound for the PM metric for the core. Localisation of the valence bands also occurs rapidly, although less markedly than in the core case, and the total number of iterations required to localise the core and valence bands separately is less than that required to localise the full occupied space (Table \ref{tab:table2}), across all the test systems. This is to be expected, given that the dimensionality of the optimisation problem is reduced by separating the core and valence bands.

Diabatic preparation is particularly effective in localising WFs from bands composed of weakly interacting AOs. The core bands and valence bands with strong ionic character, for example in MgO, require only a few optimisation steps for optimisation. Since the localised valence WFs are usually bonding in character, they are typically centred between atoms, and inherently less local than their core counterparts and require more steps for optimisaiton. Even for the core bands, using diabatic WFs as an initial preparation is vital for the robustness and accuracy of the Bloch IAO scheme.
Using a random initial unitary for core bands frequently leads to output WFs with PM values significantly smaller than those obtained with the diabatic guess, showing the encountering local maxima is a common problem without correct preparation of the WFs.

\begin{figure*}[t]
	\includegraphics[scale=0.52]{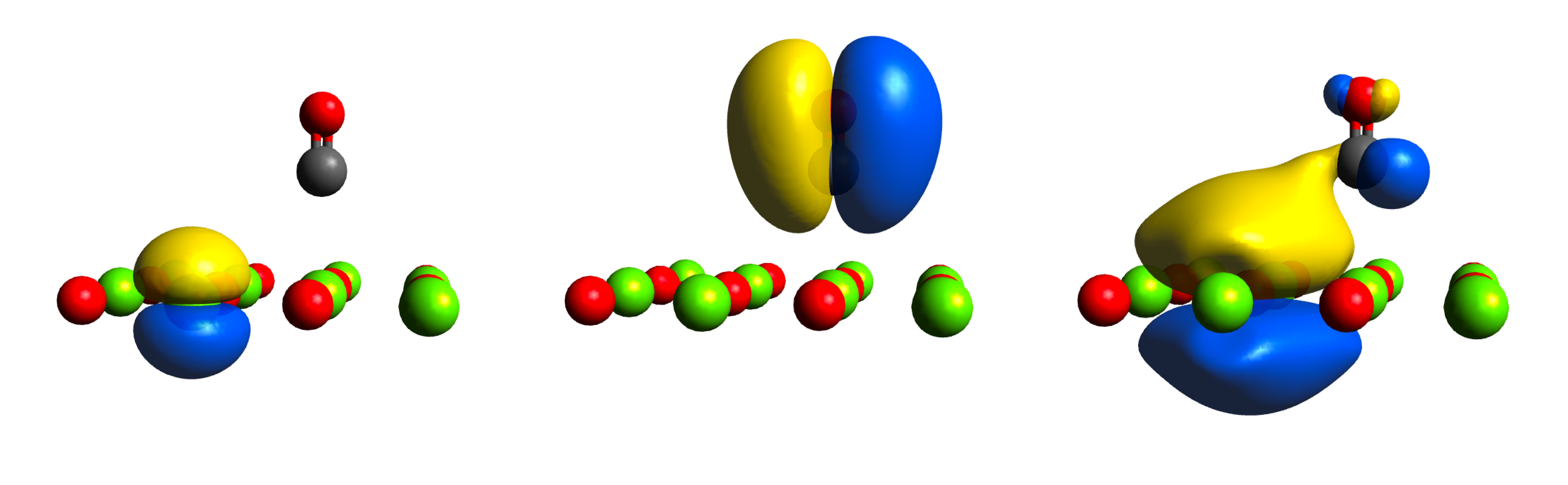}
	\caption{\label{fig:mgo_co}Bloch IAO localised Wannier functions of the MgO(001) CO adsorption system. A magnesium 2p-like orbital is shown (left), as well a $\pi$ bonding orbital on CO (centre). A WF demonstrating oxygen 3p orbital donation, on the MgO surface, to the CO $\pi^*$ orbital, is also presented (right). Only atoms in the reference cell are shown. An isosurface value of 0.01 was used.}
\end{figure*}
\subsection{Chemical intuition of Bloch IAO generated WFs}
One of the key strengths of the original IAO scheme is the clear interpretation of these molecular orbitals, and the direct connection to chemical intuition and concepts. Knizia\cite{knizia2013} demonstrated that IAOs allow robust, basis set independent, partial charges and orbital populations to be computed. This enables quantitative measures for electronegativities and oxidation states for molecules, which align with empirical understanding, to be evaluated.
In analogous fashion, we demonstrate that Bloch IAO localised WFs provide chemical understanding in periodic systems. Bloch IAO partial charges can be computed in similar fashion to moleuclar IAOs, by summing the atomic contributions across all the cells within the supercell,
\begin{equation}{\label{partialcharge}}
	q_{A}=Z_A- \sum_{\boldsymbol{R},i} Q_i^{A_{\boldsymbol{R}}^{\mathrm{IAO}}},
\end{equation}
where $Z_A$ is the atom's nuclear charge.
IAO partial charges can be computed for any periodic system, such as those listed in Table \ref{tab:table1}. However, it is also worth stressing that IAOs can be used for both periodic and molecular systems, on equal footing, meaning that they are a robust and consistent partial charge estimate for probing both material and molecular systems. This opens the possibility to investigate interesting chemical scenarios, such as systems containing interactions between material and molecules. 

The adsorption of CO onto the MgO(001) surface has been heralded as the `hydrogen molecule of surface science'\cite{h2ofsurface}, and an important case study for the theoretical understanding of heterogenous catalysis.
Obtaining an accurate adsorption energy for this system remains a highly discussed topic,
in which many quantum chemical and many body methods have been utilised\cite{comgo-vs-experiment,mgo-gagliardi,mgo-michaelides,ye2023adsorption,CO-vib-modes}, in order to achieve consensus with experimental data.
This adsorption example is an ideal case to demonstrate the ability that Bloch IAO localised WFs have to provide insight into the underlying chemistry of the system.

To model the system, a unit cell consisting of a $4\times4\times1$ slab of MgO was constructed. CO, orientated perpendicular to the surface, was positioned with a C-Mg equilibrium distance of $2.479$\r{A}\cite{mgo-gagliardi}. 
To obtain mean-field Bloch functions, a periodic DFT (PBE\cite{PBE}) calculation was conducted in the \verb:riper: module, using the pob-TZVP\cite{pob-TZVP} basis set, on a (3,3) Monkhorst-Pack mesh to sample the Brillouin zone. Bloch IAO localised WFs were then obtained.

Although higher level quantum chemical methods have been used elsewhere to attempt to accurately model the weak van der Waals interactions dominating the adsorption, we stress that our motivation is to use this system to exhibit the use of IAO WFs for chemical intuition. Since orbitals provide a zeroth order description to the motion of the electrons, and are a result from mean-field, effective one electron theories, using solely DFT to model this picture serves our purposes.

Figure \ref{fig:mgo_co} presents three Bloch IAO localised WFs of the system, all three of which align with chemical understanding. The left and centre subfigures show a localised 2p orbital, centred on Mg, and a $\pi$ bonding orbital, centred on CO, respectively. Similar WFs are observed when WFs for the surface and adsorbate are computed separately. A more interesting WF is presented on the right, where back-bonding from the nearest neighbour oxygen atom, on the MgO surface, to the $\pi^*$ orbital of carbon monoxide, is shown. Although the role of back-bonding within metal oxide adsorption has been scrutinised\cite{CO-vib-modes,mgo-co-hf-dft}
, we wish to make the point that IAO localised WFs provide direct and intuitive chemical understanding, consistent with the level of theory used to generate the original Bloch functions.

\begin{table}
	\caption{\label{tab:table4}}
	Bloch IAO partial charges of the non-interacting and equilibrium MgO(001) CO adsorption system
	\begin{ruledtabular}
		\begin{tabular}{cccccccc}
			&System&C (CO)&O (CO)& Mg (MgO)& O (MgO)\\
			\hline
			&non-interacting&0.42&-0.42&1.68&-1.68\\
			&equilibrium &0.32&-0.41&1.70\footnotemark[1]&-1.65\footnotemark[2]&&\\
		\end{tabular}
	\end{ruledtabular}
	\footnotetext[1]{The partial charge of the closest Mg atom to CO is given}
	\footnotetext[2]{The average partial charge of the four nearest neighbours to CO is given}
\end{table}

The Bloch IAO partial charges of the MgO(001) CO adsorption system, in the equilibrium geometry, were also calculated. In Table \ref{tab:table4}, we compare these partial charge values to charges obtained from separate periodic calculations of the MgO surface slab, and the CO molecule, representing a non-interacting scenario. Most significantly, a reduction in positive partial charge on the carbon of CO is observed moving from the non-interacting to equilibrium geometry, as well as a decrease in negative charge on the oxygen nearest neighbours of MgO. This can be rationalised from the back-bonding process presented in Figure \ref{fig:mgo_co}, and once again shows that quantitative measures,  such as partial charges, derived from IAO WFs, are consistent with chemical intuition, at the level of theory employed in the mean field picture.

\section{Conclusions}
We have generalised the intrinsic atomic orbital method to periodic solids. Bloch IAOs form a minimal basis which exactly represent the occupied bands, and thus alleviate the well known issue of using Mulliken charges for non-minimal basis sets. They thus enable localised Wannier functions, optimised using the Pipek-Mezey metric, to be robustly evaluated, as first introduced by J\'{o}nsson et al.\cite{Jonsson}
We outline a localisation scheme, which prepares the initial Bloch functions by diabatically transferring locality imposed at the gamma point through the Brillouin zone, before localising according to the PM metric. Clement et al.\cite{Clement} demonstrated the improved performance using l-BFGS compared to other gradient based solvers, and we confirm this. This scheme works efficiently across a range of semiconducting and insulating solids, and in particular we highlight the ability of the diabatic WFs to localise atom-centred WFs almost immediately. Using the example of CO adsorption onto MgO(001), we demonstrate that Bloch IAO localised WFs can provide chemical insight into systems, through visualisation of the WFs, and through computing measures such as partial charges. 
We expect that Bloch IAOs will provide a bridge for understanding chemical phenomena within periodic systems. Bloch IAOs are not solely restricted to LCAO methods, but can also be applied with plane wave basis sets by computing the overlap between plane waves and the minimal Gaussian AO basis\cite{Gruneis}.
In particular, we note that in the molecular setting, localised MOs using IAOs have proved popular in constructing localised occupied orbitals and domains, for use within local correlation theories\cite{pnof12}. Bloch IAOs may provide an analogous route for similar implementations within periodic systems.

\begin{acknowledgments}
Financial support from the University of Oxford and Turbomole GmbH is gratefully acknowledged.
\end{acknowledgments}

\bibliography{apssamp}

\end{document}